\documentclass[12pt,a4paper]{article}
\usepackage[cp1251]{inputenc}
\usepackage{amssymb, amsfonts, amsmath}
\usepackage[hypertex]{hyperref}

\newcommand{\rank}{{\rm rank}\,}
\newcommand{\ind}{{\rm ind}\,}
\newcommand{\dimm}{\operatorname{dim}}

\begin{document}


\title{One-loop stress tensors for scalar and spinor fields on the homogeneous spaces with $G$-invariant metrics}

\author{~A~I Breev
\thanks{Tomsk State University, Lenina 36, Tomsk 634050, Russia. E-mail:
breev@mail.tsu.ru }}

\maketitle

\begin{abstract}   
   Vacuum expectation value of the stress-energy-momentum tensor for scalar and spinor fields is obtainted on the homogeneous space with $G$-invariant metrics using the orbits of coadjoint representation of Lie group and the generalized harmonic analysis.
\end{abstract}


\section*{Introduction}
 
Quantum effects in external gravitational fields are of interest of quantum field theory.
The main object of researches in the field is the stress-energy-momentum tensor (SEMT), which is a source of gravity in Einstein's field equation. Vacuum expectation value of the SEMT gives a way to describe effects of quantum field on the space-time geometry in the framework of quantum field theory in the curved space-time \cite {Grib, Devis, Fulling}). \sloppy

In calculation of the vacuum expectations in the homogeneous and isotropic space the 
dimensional regularization \cite{ren_dim} and point-splitting \cite{ren_div} regularization are used. Moreover,  
regularization can be carried out in the vacuum expectations directly (e.g. by the $n$ - wave regularization\cite{ren_h}). 

Here, we obtain the vacuum expectations value of the SEMT of a scalar and spinor fields on the $\mathbb {R}^1\times P$ 
manifold, where $P$ is the right homogeneous space with $G$-invariant metrics and zero defect. The problem is reduced to calculations on the group $\mathbb {R}^1\times{G} $, where $G$ is the transformations group of the 
homogeneous space.

The most of the general relativity models are related to various transformations groups and corresponding homogeneous spaces.
The well knows examples of such manifolds are the De Sitter spaces \cite{desitter_1,desitter_2}. 

In calculation of the SEMT vacuum expectations we use the method of orbits of coadjoint representation of Lie groups 
(the K-orbit method) \cite{kirr1, kirr2,kirr, Constant, Souriau}. The K-orbit method involves the noncommutative algebra of symmetry operators of the wave equation together with the well known method of separation of variables \cite{shap_1,shap_2}.

\section{$G$--invariant metrics on homogeneous spaces}\label{gl1}

Let $G$ be a real connected Lie group and $L$ its Lie algebra. Let $P$ be a right $G$-space, where $G$ is a Lie 
transformation group of the homogeneous space $P$. The homogeneous space $P$ can be then represented as manifold $G/H$, where $H$ is stationary subgroup at a point $y_0\in P$ and $\mathfrak{h}$ is the Lie algebra of the group $H$, $L^{*}$ is the dual space to $L$. The left and right invariant vector fields on the Lie group $G$ are $\xi_X(x) = (L_x)_{*}X,\quad\eta_X(x)=-(R_x)_{*}X$ respectively. The linear space of smooth vector fields on the Lie group $G$ is $\mathcal{X}(G)$,
and linear space of smooth vector fields on the $P$ is $\mathcal{X}(P)$.

The principal bundle $G(P,H,\pi)$ is correspondent to the homogeneous space $P$. Let's introduce the coordinates in the trivialization domain $U\times H$ of the fibered space $G$. We obtain the local splitting of the coordinates $\{x^A\}$ on the fibered manifold in the fiber $h^{\alpha}$ and base $y^a$ coordinates, i.e. $\{x^A = (y^a,h^{\alpha})\} \, (a=1..\dimm P,\alpha=\dimm P+1..\dimm G)$. For a smooth local section $s: P \rightarrow G$ of the bundle $G$, coordinates of an arbitrary point $x\in G$ can be represented as $x=h s(y)$. For the Lie algebra $L$ we have $L = \mathfrak{h}\oplus\mathfrak{m}$, where $\mathfrak{m}$ is complement to $\mathfrak{h}$.

The vector field $\tau\in\mathcal{X}(G) $ is called \textit{projected on a homogeneous space} $P$ if 
$\pi _ {*} ([\eta_Y, \tau]) =0, Y\in\mathfrak {h}$. The projected vector fields forms a subalgebra $\mathcal{B}(G)$ in $\mathcal{X}(G)$ and $ \mathcal {X} (P) \simeq \mathcal{B}(G)/ker\{\pi _{*} \}$.

Let $\textbf{G}$ be a nondegenerate quadratic form on Lie algebra $L$, which is the metrics on the Lie group $G$
related to the unit element of $G$. With the right shifts, the right invariant metrics on the Lie group $G$ is given by:
\begin{equation}\label{left_metric}
   \langle \sigma , \tau \rangle_{x} = 
   \textbf{G} ( (R_x^{-1})_{*}\sigma , (R_x^{-1})_{*}\tau ),
   \quad \sigma,\tau\in T_x G.
\end{equation}
Then the corresponding metric tensor $\gamma_{ij}(x)$ on $G$ ($i,j=1,...,\operatorname{dim}{L}$) reads
\[
  \gamma_{ij}(x) =
  \gamma_{A B}\sigma^A_i(x)\sigma^B_j(x),
  \quad \gamma_{A B} \equiv \textbf{G}(e_A, e_B),
\]
where $\{e_A\}$ is a given basis of the algebra $L$ and $\sigma^B(x)$ are components of a basis of right invariant 1-forms: $\sigma^{B}(x)\equiv (R_x)^{*}e^B$. The dual basis of the coalgebra $L^{*}$ is $\{e^B\}$, i.e.$\langle e_A,e^B\rangle = \delta_A^B$, ($A,B=1,...,\operatorname{dim}{L}$).

Consider $G$-invarint metric on the homogeneous space $P$ corresponding to right invariant metrics 
on the Lie group $G$. Let the 1-form $\textbf{G}$ satisfies the condition
\begin{equation}\nonumber
  \textbf{G}([X,Y],Z)+\textbf{G}(Y,[X,Z])=0,\quad X\in \mathfrak{h}, Y,Z\in L. \label{conds_2}
\end{equation}
Define a 1-form $\textbf{B}$ on the Lie algebra $L$ as
\begin{gather}\nonumber
  \ker \textbf{B} = \mathfrak{h}, \label{conds_1} \\ \nonumber
  \textbf{B}(X,Y) = \textbf{G}(X,Y), \quad X,Y\in\mathfrak{m}.
\end{gather}
The G-invariant metrics on the homogeneous space $P$ has the form
\begin{equation}\label{metric_P}
   \langle  \pi_{*}(\tau), \pi_{*}(\textit{v})  \rangle_{\pi(x)} \equiv
   \textbf{B}( (R_x^{-1})_{*} \sigma,(R_x^{-1})_{*} \tau ),\quad \textit{v},\tau\in \mathcal{B}(G).
\end{equation}
The Christoffel symbols related to the metrics (\ref{metric_P}) have the form
\begin{equation}\label{gamma_P_IJK}     
   \Gamma_{jk}^i(y) = \Gamma_{b c}^a \sigma^b_j(x)\sigma^c_k(x)\eta_a^i(x) - 
   \sigma^b_j(x)\eta_{b,k}^i - C_{b\alpha}^a\sigma^b_j(x)\sigma_k^{\alpha}(x)\eta_a^i(x),
\end{equation}   
where $\Gamma_{b c}^a$ is
\begin{equation}\label{gamma_P}
  \Gamma_{b c}^a = -\frac{1}{2}C_{bc}^a - \frac{1}{2}G^{ad}
  \left[G_{e c}C_{b d}^e + G_{e b} C_{c d}^e\right].
\end{equation}
The correspondent the Riemann tensor on the homogeneous space $P$ is:
\begin{gather}\nonumber
   R^i_{j k l }(y) = R^a_{b c d}\sigma^b_i(x)\sigma^c_k(x)\sigma^d_l(x)\eta_a^i(x),\\
   R^a_{b c d} = \Gamma^a_{\bar{d}d}\Gamma^{\bar{d}}_{b c} - \Gamma^a_{\bar{d}c}\Gamma^{\bar{d}}_{b d} +
                 C^{\bar{c}}_{c d}\Gamma^a_{b \bar{c}}.
\end{gather}

\section{Harmonic analysis on Lie groups and homogeneous spaces}

We can set a function from $\hat\varphi\in C^{\infty}(P)$ into correspondence with any function $\varphi\in C^{\infty}(G)$, but this correspondence holds only in a subspace $\mathcal{F} \subset  C^{\infty}(G)$, which can be selected by a special condition. Define the space $\mathcal{F}$ as
\[
  \mathcal{F} \equiv \{ \varphi\in C^{\infty}(G) \,  \mid \, \eta_{X}\varphi(x)=0 ,\ X\in\mathfrak{h},
  \, x\in G\}.
\]
Here $\mathcal{F}$ is a space of functions on $G$ that are constant on the fibers of the bundle $G(P,H,\pi)$. 

The Lie group $G$ acts on the dual spaces $L^* = T_e^{*}G$ by the coadjoint representation
$Ad^*: G\times L^* \rightarrow L^*, (x,f)\rightarrow Ad_x^* f$, where $\mathcal{O}_{\lambda}\approx L^{*}/G^{\lambda}$,
$G^{\lambda} = \{ Ad_{x}^{*}\lambda=\lambda \mid x\in G \}$ is the orbit of the coadjoint representation of $G$ passing through the linear functional $\lambda\in L^{*}$, and $\omega_{\lambda}$ is the Kirillov symplectic form
\begin{equation}\label{kirr_form}
   \omega_{\lambda}(a,b)=\langle\lambda,[X,Y]\rangle,\quad
   a = ad_{X}^*\lambda,b=ad_{Y}^*\lambda \in T_{\lambda}\mathcal{O}_{\lambda},\quad X,Y\in L.
\end{equation}
According to the Darboux theorem the local canonical coordinates (the Darboux coordinates) exist on the orbit $\mathcal{O}_{\lambda}$ in which the 2-form $\omega_{\lambda}$ is $\omega_{\lambda} = dp_a \wedge dq^a, a = 1..\operatorname*{dim}\mathcal{O}_{\lambda}/2$. The domain $Q$ of the variables $q$ and the domain $P$ of the variables $p$ are a Lagrangian submanifolds of the orbit $\mathcal{O}_{\lambda}$.

Smooth functions $K_{\mu}: L^{*} \rightarrow \mathbb{R}^1$, $\mu=1,...,\operatorname*{ind} L$ are called  \textit{Casimir functions} if $\{ K_{\mu},\varphi \} \equiv 0$ for every $\varphi\in C^{\infty}(L^{*})$. In other words, the Casimir functions
are those from the center of $C^{\infty}(L^{*})$ viewed as the Lie algebra w.r.t. the Poisson bracket. 
The Casimir functions are locally invariant under the coadjoint representation, $K_{\mu}(Ad_x^* f)=K_{\mu}(f),x\in G$. The number $\operatorname*{ind}L$ of independed Casimir functions is called the \textit{index} of the Lie algebra $L$.

The action of adjoint representation of the Lie group defines a fibration of the dual space $L^*$ into 
even-dimensional orbits (the K-orbits). The dimension of a K-orbit is $\dimm L - \ind L - 2 k$, where $k=0\dots (\dimm L - \ind L)/2$.The space $L^{*}$ consists of the union of connected djsjoint invariant algebraic surfaces $M_{(s)}$, where $M_{(s)}$ is an space representable as a union of orbits of the same dimension $\dimm G - \ind L - 2 s$. The functions $K^{(s)}_{\mu}(f)$ that are nonconstant on $M_{(s)}$ are called the $(s)$-\textit{type Casimir functions} if they commute with any function on $M_{(s)}$. The orbits from $M_{(s)}$ are called the $(s)$-\textit{type K-orbits}. 

Denote by $r_{(s)}$ number independed Casimir functions of type $(s)$. We say that the K-orbit is nondegenerate if $s=0$. Denote by $F_{\alpha}^{(s)}(f), \alpha=1..\operatorname{dim}L-r_{(s)}$ an independent set of functions defining a surface $M_{(s)}$.

It is easy to see that the transition to the canonical Darboux coordinates $f\rightarrow (p,q)$ means to find a set
of analytic functions of the variables $(p,q): f_X=f_X(q,p,\lambda), X\in L$ that satisfy the following conditions:
\begin{gather}\label{can_f}
  \{f_X,f_Y\}^{Lie} = f_{[X,Y]}, \quad f_X(0,0,\lambda) = \lambda(X), \quad
  X,Y\in L,\\
  F_{\alpha}^{(s)}(f) = 0,\ (\alpha = 1..\operatorname{dim}L-r_{(s)}).  
\end{gather}
Let the transition to the canonical coordinates be linear on $p$, i.e.
\begin{equation}\label{f_X}
   f_X (q,p,\lambda) = \alpha^a_{X}(q)p_a + \chi_X(q,\lambda), \quad
   X\in L,\quad a = 1..\operatorname*{dim}\mathcal{O}_{\lambda}/2.
\end{equation}
The subalgebra $\mathfrak{n} \subset L_{\mathbb{C}}$, such that 
$$
  \operatorname*{dim}\mathfrak{n} = \operatorname*{dim} L - \operatorname*{dim}\mathcal{O}_{\lambda}/2,\quad
  \langle \lambda,[\textbf{n},\textbf{n}]\rangle=0,
$$
is called \textit{polarization} of the linear functional $\lambda$. 
Polarization is shown to exist for arbitrary Lie algebra and for any nondegenerate functional $\lambda$ \cite{Bar2}.

Linear transition to the canonical coordinates (\ref{f_X}) exists only when for a linear functional $\lambda$
the polarization does exist.

Define the special infinite-dimensional irreducible representation of the Lie algebra $L$ 
on the space of smooth functions $L(Q,\textbf{n},\lambda)$ which we call
$\lambda$-representation. Operators of $\lambda$--representation $l_X(q,\lambda)$ are constructed as follows from
the quantized canonical of transition functions $l_X(q,\lambda) \equiv i f_X(q,-i\partial_q,\lambda)$.

The scalar product on the manifold $Q$ with the measure $d\mu(q)$ is given by
\begin{equation}\label{sc_Q}
  (\psi_1,\psi_2) = \int_Q \overline{\psi_1(q)}\psi_2(q) d\mu(q).
\end{equation}
We introduce a ``quantum shift'' in the operators $l$ as $\lambda\rightarrow\lambda+i\beta$, where
\begin{equation}\label{beta_kv}
   \beta(X) = \frac{1}{2}\left[ tr(ad_X) - tr(ad_X|_{\mathfrak{l}}) \right],\quad
   \mathfrak{l} = \mathfrak{n} + \bar{\mathfrak{n}}.
\end{equation}
Then operators $l_X$ are anti-Hermitean with respect to the scalar product (\label{sc_Q}).

Let $T^{\lambda}$ be the lift of the $\lambda$-representation to a local representation of the Lie
group $G$:
\begin{gather}\label{Tj}
   T^{\lambda}(x)\varphi(q)=\int D_{q\overline q'}^{\lambda}(x)\varphi(q') d\mu(q'),\\ \nonumber
   \frac{\partial}{\partial t}|_{t=0}T^{\lambda}(\exp(t X))\varphi(q) = l_{X}(q,\lambda)\varphi(q),
\end{gather}
where $\varphi\in C^{\infty}(Q)$. The condition $T^{\lambda}(x_1)T^{\lambda}(x_2)=T^{\lambda}(x_1 x_2)$ 
implies a relation for "matrix elements" of the representation $T^{\lambda}$:
\begin{equation}\label{DD}
   D_{q\overline q'}^{\lambda}(x_1 x_2) = \int D_{q\overline q''}^{\lambda}(x_1)
                                      D_{q''\overline q'}^{\lambda}(x_2) d\mu(q'').
\end{equation}
The generalized functions $D_{q \overline q'}^{\lambda} (x)$  satisfy the overdetermined system of equations
\begin{equation}\label{D_system}
   [\eta_X(x)+l_X(q,\lambda)] D^{\lambda}_{q\overline q'}(x)=0, \quad
   [\xi_X(x)-\overline{l_X^{\dag}(q',\lambda)}] D^{\lambda}_{q\overline q'}(x)=0.
\end{equation}

System (\ref{D_system}) is consistent but global solutions exist only for integer-valued orbits \cite{kirr,Sh_2}
such that
\begin{equation}\nonumber
   \frac{1}{2\pi}\int_{\gamma\in H_2 (\mathcal{O_{\lambda}})} \omega_{\lambda} =
   n_{\gamma}\in\mathbb{Z},
\end{equation}
where $\omega_{\lambda}$ is the Kirillov form on the K-orbit.

The distributions $D_{q \overline {q'}}^{\lambda}(x)$ obey the completeness and orthogonality relations
   \begin{align}
     \label{DD_1}  
     & \int_{G} \overline{\tilde{D}^{\tilde\lambda}_{\tilde q \overline{\tilde q}'}}(x)
                D^{\lambda}_{q\overline{q}'}(x) d\mu(x) = 
                \delta(q,\overline{\tilde{q}})
                \delta(\tilde{q}',\overline{q}')\delta(\tilde{\lambda},\lambda), \\
     \label{DD_2}             
     & \int_{Q\times Q\times J}\overline{\tilde{D}^{\lambda}_{q\overline{q}'}(\tilde{x})}
                               D^{\lambda}_{q\overline{q}'}(x)d\mu(q)d\mu(q')d\mu(\lambda)=
       \delta(\tilde{x},x).                           
   \end{align}    
For any function $\phi(x)$ from the dense nuclear subspace
$$
  L_{(s)}=\{\phi\in L_2(G,d x) | F_{\alpha}^{(s)}(\xi_X(x)\phi(x)) = 0, X\in L \}
$$ 
the direct and inverse Fourier transformation are
\begin{align}
  \label{F_1}
  & \psi( q,q',\lambda) = \int \phi(x) \overline{D^{\lambda}_{q\overline q'}}(x^{-1}) d\mu(x),\\
  \label{F_2}
  & \phi(x) = \int \psi(q,q',\lambda) D^{\lambda}_{q\overline q'}(x^{-1}) d\mu(q)d\mu(q')d\mu(\lambda),
\end{align}
where $d\mu(\lambda)$ is the spectral measure of the Casimir operators $K_{\mu}(\eta)$. 

If a K-orbit $\mathcal{O}_{\lambda}$ is nondegenerate then transformations (\ref{F_1})-(\ref{F_2}) act on
$L_2(G, d x)$.

Note that if the functions $\varphi(x)$ and $\psi(q,q',\lambda)$ are connected by the transformations
 (\ref{F_1})-(\ref{F_2}) then the same transformations connect the operators 
\begin{equation}\nonumber
    \eta_X(x) \Leftrightarrow l_X(q',\lambda),\ 
    \xi_X(x) \Leftrightarrow \overline{l^{\dag}_X(q,\lambda)}.
\end{equation}
For example, the differential operator acting in the space $Q$ with a smaller number of independent variables corresponds
to the Laplace operator on the Lie group (enveloping algebra element of the left- and right-invariant vector fields).

Define the functional subspaces $L_{(s)}(P) \subset L_2(P,d\mu(y))$ as
\[
  L_{(s)}(P) = L_{(s)} \cap \mathcal{F}  = 
               \{ \varphi \in L_2(P,d\mu(y)) | F^{(s)}(\tilde X)\varphi = 0 \},
\] 
where $\tilde X$ are generators of a transformation group $G$ acting on a homogeneous space $P$. In  \cite{Shir_pr}
it is shown that $L_{(s)}(P) = \emptyset, s < s_P$ and
\begin{equation}\label{cond_22}
  F^{(s_P)}(\tilde X) \equiv 0,\quad 
  \tilde{K}^{(s_P)}_{\mu}(i \tilde X) \equiv 0,\quad \mu=1,...,r_{s_P},
\end{equation} 
where value $s_P$ is degeneracy degree of the homogeneous space $P$ (see \cite{Shir_pr}); 
$\tilde{K}^{(s_P)}_{\mu}(f)$ are the $(s_P)$--type Casimir functions, such as $\tilde{K}^{(s_P)}_{\mu}|_{\mathfrak{h}^{\perp}}=0$. From (\ref{cond_22}) it follows that $L_{(s_P)}(P) = L_2(P,d\mu(y))$.

The $\lambda$-representation of the Lie algebra $L$ corresponds to the homogeneous space if
\begin{equation}\label{cond_22_II}
  \tilde{K}_{\mu}^{(s_P)}(\lambda(j)) = 0,\mu = 1,\dots,r_{s_P}.
\end{equation}

For the $s_P$--type $\lambda$-representation of the algebra $L$ the set of generalized functions $D^{\lambda}_{q \overline{q}'}(x)$ in the functional space $L_{(s_P)}$ obeys the completeness and orthogonality relations.
 The transformation (\ref{F_1})-(\ref{F_2}) holds for homogeneous space $P$ if 
\begin{equation}\label{cond_P}
  l_{\alpha}(q',\lambda)\psi(q,q',\lambda) = 0,\quad  \alpha=1..\operatorname*{dim} H.
\end{equation}
Every homogeneous space $P$ is characterized by a number
\[
  d_P = \frac{1}{2} \rm{sup} \, \rank \langle \lambda,[L,L] \rangle - 
                    \rm{sup} \, \rank \langle \lambda,[L,\mathfrak{h}] \rangle
,\quad \lambda\in \mathfrak{h}^{\perp},
\]
which is called \textit{defect} of $P$. 

If $d_P = 0$, then the algebra of invariant differential
operators is commutative and the system (\ref{cond_P}) always has a solution. If $d_P > 0$ then  the algebra of invariant operators is noncommutative and the construction of harmonious analysis would demand introduction of special methods \cite{Shir_pr}. 

\section{Integration of the Klein-Gordon and Dirac equations}

Consider the $(n+1)$-dimensional manifold $M^{n+1}=\mathbb{R}^1\times P^{n}$ where an
element $y'\in P$ is denoted by $(y,t), y\in P, \quad t\in\mathbb{R}$. Components of $G$-invariant metrics 
$G_{\tilde A\tilde B}$ on the $L^{n+1} = \mathbb{R}^1\times L$ are $G_{0 0} = 1,\quad G_{0 a} = 0,\quad G_{a b} = -\gamma_{a b}$, $A',B',...=0,1,\dots,n$. $\gamma_{a b}$ are components of $G$-invariant metric on the $P^{n}$. 

The Klein-Gordon equation with conformal connection $\zeta = (n-1)/(4 n)$ on the manifold $M$ has the form
\begin{equation}\label{clein_gordon}
   (\partial^2_t - \Delta_M + \zeta R + m^2)\varphi (x,t) = 0, 
\end{equation}
where $\Delta_M = \gamma^{ab}[ \eta_a\eta_b - Sp(ad_a)\eta_b ]|_{\mathcal{F}}$ is the Laplace operator on the  $P$; $R$ is the scalar curvature of the manifold $M$. The indefinite scalar product associated with Eq. (\ref{clein_gordon}) is
\begin{equation}\label{sc_inner_product}
   ( \varphi_1, \varphi_2 ) = -i\int \varphi_1\overleftrightarrow{\partial_{t}}\varphi_2 d\mu(y),
\end{equation}
where $\varphi_1, \varphi_2$ are two solutions of (\ref{clein_gordon}).

We now construct the basic of solutions $\varphi_{\sigma}(y,t)$ of Eq. (\ref{clein_gordon}) enumerated by the 
index $\sigma$ and which are normalized with respect to the inner product (\ref{sc_inner_product}).
We seek solution of the form $\varphi(y,t) = f(t)F(y)$. Substituting,
\begin{equation}\label{ft}
  f_{\Lambda}(t) = \frac{1}{\sqrt{2\omega}} e^{-i\omega t},
                   \quad \omega^2 = \Lambda^2 + \zeta R + m^2,
\end{equation}
where $\Lambda$ is a separation constant. For the function $F(y)$, we have equation:
\begin{equation} \label{clein_fok}
   -\Delta_M F_{\Lambda}(y) = \Lambda^2 F_{\Lambda}(y).
\end{equation}
Using transformation (\ref{F_1}), we reduce Eq. (\ref{clein_fok}) to the equation on the K-orbit. In this case,
the Laplace operator $\Delta_M$ is transformed into a constant $H(\lambda)$.
The functions
\begin{gather}\label{cl_g_F}
   F_{\sigma}(y) = \int \psi(q',\lambda) D^{\lambda}_{q \overline{q'}}(x^{-1}) d\mu(q'),  \quad \sigma=(q,\lambda),
\end{gather}
form a basic of solutions of Eq. (\ref{clein_fok}). The function $\psi(q',\lambda)$ is a solution of Eq. (\ref{cond_P})
and satisfy the orthogonality condition
\begin{equation}\label{ort_psi}
   \int_Q{ \psi_{\Lambda}^{\dag}(q,\lambda)
          \psi_{\Lambda'}(q,\lambda)} d\mu(q) = 
   \delta(\Lambda,\Lambda').
\end{equation}
This relationship is the normalization condition of the function $\psi_{\Lambda}(q',\lambda)$ and is the definion of 
the measure of the eigenvalues $\Lambda$.

Therefore, the basic of solutions for Eq. (\ref{clein_gordon}) has the form 
\begin{equation}\label{basic_eq}
\Psi_{\sigma}(x,t) = f_{\Lambda}(t)F_{\sigma}(y),
\end{equation}
where $f_{\Lambda}(t)$ is determined by (\ref{ft}), and $F_{\sigma}(y)$ is determined by (\ref{cl_g_F}).

In the general case, the Dirac equation on the Riemannian space for the particle with mass $m$ and spinor $\Psi(y)$
has the form
\begin{equation}\label{dirac1}
   \left(i\gamma^i(y)\vec{\nabla}_i -m \right)\Psi(y) = 0,
\end{equation}
where $\vec{\nabla}_i \equiv \nabla_i + \Gamma_i(y)$; $\nabla_i$ is the covariant derivative,
$\Gamma_i(y) = -1/4 \gamma_{i;k}(y)\gamma^i(y)$ is the spinor connection, i.e. $[\vec{\nabla}_i,\gamma_j(y)]=0$, 
$\operatorname*{Tr} \Gamma_k(y) = 0$  ($i,j,k=0,...,n$).
The Dirac matrices on the Riemannian space are determined as an arbitrary but fixed solutions of the system 
$\{\gamma_i(y),\gamma_j(y)\}=2 g_{ij}(y) E_4$, where $E_4$ is the identity matrix, $g_{ij}(y)$ is the metric tensor.

The Dirac equation (\ref{dirac1}) on $M$ assumes the form
\begin{equation}\label{dirac}
   \left(i\gamma^0\partial_t  + i\gamma^a[\eta_a(y,h)+\Gamma_a] - m \right)\Psi(y) = 0,\quad
   \Gamma_a = -\frac{1}{4}\Gamma^c_{ba}\gamma^b\gamma_c.
\end{equation}
The constant matrices $\gamma^0$ and $\gamma^a = \eta^a_i(y)\gamma^i(y)$ satisfy the relationship 
$\{\gamma_a,\gamma_b\}=2 \gamma_{ab} E_4$ and form the representation of the Clifford algebra constructed with respect to the 
bilinear form $\gamma_{ab}$. If $\gamma_{ab}=\operatorname*{diag}(1,-1,-1,-1)$, we obtain the usual Dirac 
$\gamma$-matrices.

Let us find a solution basis of the Dirac equation (\ref{dirac}). We seek solution of the form
\begin{equation}\label{razd}
   \Psi(y,t) = f(t)F(y).
\end{equation}
Substituting (\ref{razd}) into (\ref{dirac}), we get
\begin{equation}\label{ft2}
   f_{\Lambda}(t) = exp(-i\omega t),\quad \omega^2 = \Lambda^2.
\end{equation}
The normalization is taken in the form $|f_{\Lambda}(t)|^2 = 1$. The equation for the spinor $F(y)$ is
\begin{equation}\label{dirac_G}
   D(\eta) F_{\Lambda s}(y) = 0,\quad D(\eta) = \gamma^a(\eta_a(x) + \Gamma_a) - i\Lambda\gamma^0 + i m,
\end{equation}
where $s$ is a spin index. We seek for a solution of Eq. (\ref{dirac_G}) in the form of Eq. (\ref{F_2}). Thus for the
spinor $\psi_{\Lambda s}(q, q',\lambda)$ we have
\begin{equation} \nonumber
  -D(l(q',\lambda))\psi_{\Lambda s}(q, q', \lambda) = 
  \Lambda^2 \psi_{\Lambda s}( q,q',\lambda).
\end{equation}
The spinors
\begin{equation}\label{find_F}
   F_{\sigma}(y) = \int \psi_{\Lambda s}(q',\lambda)D_{q \overline q'}^{\lambda}(y,h)d\mu(q'),  
   \quad \sigma=(q,\lambda,\Lambda, s)
\end{equation}
form a solution basic of Eq.(\ref{dirac_G}). The functions $\psi_{\Lambda s}(q',\lambda)$ can be found from the 
equation in the K-orbit
\begin{equation}\label{clein_fok_K}
    - D(l(q',\lambda)) \psi_{\Lambda}(q',\lambda) = 
    \Lambda^2 \psi_{\Lambda}(q',\lambda)
\end{equation}
The spinors $\psi_{\Lambda s}(q',\lambda)$ satisfy the orthogonality condition
\begin{equation}\label{ort_psi}
   \int_Q{ \psi_{\Lambda s}^{\dag}(q',\lambda)
          \psi_{\Lambda' s'}(q',\lambda)} d\mu(q') = 
   \delta(\Lambda,\Lambda')\delta_{s s'},
\end{equation}
where $\psi^{\dag}$ is the Hermitian conjugation of the spinor. This relationship specifies the normalization condition
of the spinor $\psi_{\Lambda s}(q',\lambda)$ and defines the measure in terms of the eigenvalues $\Lambda$.
 
Thus, the basic of solutions of Eq. (\ref{dirac}) on the manifold $M^{n+1}$ numbered by the subscript $\sigma$ has the form
\begin{equation}\label{basic_eq}
\Psi_{\sigma}(x) = f_{\Lambda}(t)F_{\sigma}(y)
\end{equation}
where $f_{\Lambda}(t)$ is determined by Eq. (\ref{ft2}) and $F_{\sigma}(x)$ is expressed through the functions
$\psi_{\Lambda s}(q',\lambda)$ according to formula (\ref{find_F}).  

\section{Stress tensor for scalar and spinor fields}

Consider the stress tensor for a scalar field on the homogeneous space $M^{n+1}$. The stress-tensor for the slalar field can be obtained by variation of the action of scalar field with respect to the metric \cite{Grib}. 
The local components ${T}_{ij}$ of stress tensor on the $M^{n+1}$ are 
\begin{align} \label{tensor_ep}
T_{ij}\{\overline{\varphi}, \varphi \} = 
           & (1-2\zeta)\overline{\varphi_{(,i}}\varphi_{,j)} +
                \left(2\zeta-\frac{1}{2}\right)g_{ij}g^{kl}\overline{\varphi_{,k}}\varphi_{,l}-
           \zeta [(\Delta_i\Delta_j\overline{\varphi})\varphi + 
              \overline{\varphi}(\Delta_i\Delta_j\varphi)]- \\ \nonumber
         & \left[\zeta R_{ij} + \left(2\zeta-\frac{1}{2}\right)g_{ij}(m^2+
           \zeta R)\right]|\varphi|^2.
\end{align}
Consider components
\begin{equation}\label{kvt}
  {T}_{\tilde{A} \tilde{B}}(x,t) \equiv {T}_{ij}(y,t) \eta_{\tilde{A}}^i(y',h) \eta_{\tilde{B}}^j(y',h).   
\end{equation} 
Since
$
  \eta_{\tilde{A}}^i\sigma^{\tilde{A}}_j=\delta^i_j,\quad 
  \eta_{\tilde{A}}^i\sigma^{\tilde{B}}_i=\delta_{\tilde{A}}^{\tilde{B}},
$ 
always it is possible to proceed from components (\ref {kvt}) to local coordinates and back.
If $P$ is Lie group ($\operatorname*{dim}H=0$) then components (\ref{kvt}) are tetrad components of tensor field $T$.

For the ``quasi-tetrad`` components of the SEMT we obtain
\begin{align}\label{EMT_P}
   & T_{ \tilde{A} \tilde{B}}\{ \overline{\varphi},\varphi \}(x,t) =       
   \left(1-2{\zeta}\right)\eta_{\tilde{A}}(y',h)
   \overline{\varphi(y')}\eta_{\tilde{B}}(y',h)\varphi(y') +\\ \nonumber
   &+ \left(2\zeta-\frac{1}{2}\right)
   G_{\tilde{A} \tilde{B}}G^{\tilde{C} \tilde{D}}
   \eta_{\tilde{C}}(y',t)\overline{\varphi(y')}\eta_{\tilde{D}}\varphi(y')- \\ 
           \nonumber
           & -\zeta \left[ \overline{\varphi(y')}(\eta_{\tilde{A}}(y',h)\eta_{\tilde{B}}(y',h)\varphi(y')) +
(\eta_{\tilde{A}}(y',h)\eta_{\tilde{B}}(y',h)\overline{\varphi(y')})\varphi(y')\right]-\\ \nonumber
   &-\zeta\left( \overline{\varphi(y')}\eta_{C}(y',h)\varphi(y') + 
                \eta_{C}(y',h)\overline{\varphi(y')}\varphi(y') \right) 
   \Gamma_{\tilde{B} \tilde{A}}^c - \\ \nonumber
   &-\left( \zeta R_{\tilde{A} \tilde{B}} + 
    \left(2\zeta-\frac{1}{2}\right)G_{\tilde{A} \tilde{B}} (m^2+\zeta R) \right)  
   |\varphi(y')|^2.
\end{align}
The system is quantized in the canonical quantization scheme by treating the field $\phi$ as an operator and imposing
commutation relation
\begin{equation}\nonumber
   \left[ \hat\varphi(y,t), \hat\eta_0 \varphi(y,t') \right] = i \delta(y,y').
\end{equation}
The field modes (\ref{basic_eq}) form a complete orthonormal basic with scalar product (\ref{sc_Q}), so operator
field function may be expanded as
\begin{equation}\nonumber
   \hat \phi(y,t) = \int d\mu(\sigma)
   \left[ \varphi(y,t) \hat a_{\sigma} + 
  \overline{\varphi(y,t)}\hat a^{\dag}_{\sigma} \right],
\end{equation}
where $\hat a^{\dag}_{\sigma}$ is the generation operator, $\hat a_{\sigma}$ -- is the generation operator; The averaged stress tensors are determined from the formula
\begin{equation}\label{va_def}
  \langle \hat T_{ij} \rangle = 
   \int T_{ij} \{ \overline{\varphi}_{\sigma} \varphi_{\sigma} \}
                            d\mu(\sigma),\quad X,Y\in \mathbb{R}^1\times L.
\end{equation}
Let's substitude (\ref{basic_eq}) in (\ref{va_def}) and integrate over $q$ taking into account special properties of
functions $D^{\lambda}_{q\overline q'}(x)$, we will get expression 
\begin{gather}\label{T00_vac}
  \langle  \hat T_{0 0} \rangle =-\frac{1}{2}\int \omega
   | \psi(q,\lambda) |^2 d\mu(q)d\mu(\lambda), \\  \label{T0a_vac}        
  \langle  \hat T_{0 a} \rangle = 
  \frac{i}{2}\int 
  \overline{\psi(q,\lambda)} \left(l_a(q,\lambda) \psi(q,\lambda)\right)
   d\mu(q)d\mu(\lambda),\\  \label{TAB_vac}         
  \langle  \hat T_{a b} \rangle = 
  \int  \frac{1}{2\omega}\overline{\psi(q,\lambda)}\bigg{(} 
  \{ l_{a},l_{b} \}_{+} - \tilde\zeta R_{a b} \bigg{)} \psi(q,\lambda)d\mu(q)d\mu(\lambda),
\end{gather}
where $l_{a}=l_{a}(q,\lambda)$ is operators of $\lambda$ representation of type $s_P$ corresponding to homogeneous space (see (\ref{cond_22_II})); $\psi(q,\lambda)$ -- solution of Eq. (\ref{cond_P}); 

The SEMT for a spinor field in the four-dimensional Riemannian space is \cite{Devis}:
\begin{equation}\nonumber
   T^{1/2}_{ij}\{ \overline{\Psi},\Psi \} = \frac{i}{2}\left[
                     \overline{\Psi}\gamma_{(i}\stackrel{\rightarrow}{\nabla_{j)}}\Psi-
                     \overline{\stackrel{\rightarrow}{\nabla_{(i}}\Psi}\gamma_{j)}\Psi
   \right], 
\end{equation} 
where the designations $\overline\Psi \equiv \Psi^{\dag}\gamma_0$ have been introduced, the overbar 
denotes the Dirac conjugation $\overline{\Psi(y)}\equiv \Psi^{\dag}(y)\gamma_{0}$, where $\Psi(y)$ are the field functions (spinors). 

The quantization of spin $1/2$ fermion field proceeds in close analogy. We can expand the field $\hat \Psi(y,t)$ as 
\begin{equation}\label{hat_Psi}
   \hat \Psi(y,t) = \int {\left(\psi^{(+)}_{\sigma}(y,t) \hat a_{\sigma} + 
                        \psi^{(-)}_{\sigma}(y,t) \hat b^{\dag}_{\sigma}\right)}d\mu(\sigma).
\end{equation}
with noncommutative relationships:
\begin{equation}\nonumber
   \{\hat \Psi(y,t), i\hat \Psi^{\dag}(\tilde{y},t)\} = i\delta(y,\tilde{y}),\quad
   \{\hat \Psi(y,t), i\hat \Psi(\tilde{y},t)\} = \{\hat \Psi^{\dag}(y,t), i\hat \Psi^{\dag}(\tilde{y},t)\} = 0. 
\end{equation}
In formula (\ref{hat_Psi}), the following designations have been introduced:  $\psi^{(\pm)}_{\sigma}(y,t)$ is the full set
of positive- and negative-frequency solutions of the Dirac equations, $\sigma$ in the set of quantum numbers, 
$\hat a_{\sigma}$ is the electron annihilation operator, and $\hat b^{\dag}_{\sigma}$ is the positron generation operator.

We now write the ''quasi-tetrad'' components of operator of the energy-momentum tensor for the spinor field:
\begin{align}
    \nonumber
     \hat T^{1/2}_{\tilde{A}\tilde{B}}\{ \overline{\Psi},\Psi \} = \frac{i}{2}\left[
                    (\hat{\overline{\Psi}}\gamma_{(\tilde{A}}\xi_{\tilde{B})}\hat{\Psi}-
                     \xi_{(\tilde{A}}\hat{\overline{\Psi}}\gamma_{\tilde{B})}\hat{\Psi})+
                    (\hat{\overline{\Psi}}\gamma_{(\tilde{A}}\Gamma_{\tilde{B})}\hat{\Psi}-
                     \overline{(\Gamma_{(\tilde{A}}\hat{\Psi})}\gamma_{\tilde{B})}\hat{\Psi})  
                    \right].
\end{align} 
The expectation value of the SEMT are determined from the formula
\begin{equation}\label{T_vac}
   \langle \hat T_{\tilde{A}\tilde{B}} \rangle = \int \hat T^{1/2}_{\tilde{A}\tilde{B}}\{\overline\Psi_{\Lambda}^{(-)},
                                         \Psi_{\Lambda}^{(-)}\} d\mu(\Lambda).
\end{equation}
Substituting expression (\ref{ft}) into Eq. (\ref{T_vac}) for $f_{\Lambda}(t)$, we obtain the solution of the 
Dirac equation in the form of Eq. (\ref{find_F}). We have:
\begin{align}
   \label{T00_vac_sp}
   & \langle \hat{T}_{00} \rangle = 
   -\int \Lambda\psi_{\Lambda s}^{(-)\dag}(q',\lambda)
                           \psi_{\Lambda s}^{(-)}(q',\lambda)
                                          d\mu(q')d\mu(\lambda)d\mu(\Lambda)d\mu(s),
\end{align}
\begin{align}
   \label{T0a_vac_sp}                                       
   \langle \hat{T}_{0a} \rangle = & -\frac{1}{2}\int
   \Lambda
   \overline{\psi_{\Lambda s}^{(-)}(q',\lambda)}
    \hat\gamma_a\psi_{\Lambda s}^{(-)}(q',\lambda)
                          d\mu(q')d\mu(\lambda)d\mu(\Lambda)d\mu(s) + \\ \nonumber
                       &  \frac{i}{2}\int
                          \psi_{\Lambda s}^{(-)\dag}(q',\lambda)
 l_a(q',\lambda) \psi_{\Lambda s}^{(-)}(q',\lambda)
                          d\mu(q')d\mu(\lambda)d\mu(\Lambda)d\mu(s) + \\ \nonumber
                       &  \frac{i}{4}\int\left(
                          \psi_{\Lambda s}^{(-)\dag}(q',\lambda)\Gamma_A
 \psi_{\Lambda s}^{(-)}(q',\lambda) -
                          \overline{(\Gamma_a\psi_{\Lambda s}^{(-)}(q',\lambda))}
        \psi_{\Lambda s}^{(-)}(q',\lambda)
                          \right)\times\\ \nonumber
  & \rule{1cm}{0pt}\times d\mu(q')d\mu(\lambda)d\mu(\Lambda)d\mu(s),
\end{align}
\begin{align}
   \label{Tab_vac_sp}                       
  \langle \hat{T}_{ab} \rangle = & i\int
                          \overline{\psi_{\Lambda s}^{(-)}(q',\lambda)}
        \hat\gamma_{(a} l_{b)}(q',\lambda)
                          \psi_{\Lambda s}^{(-)}(q',\lambda)\\ \nonumber
 & \rule{1cm}{0pt}\times d\mu(q')d\mu(\lambda)d\mu(\Lambda)d\mu(s) + \\  \nonumber
                       & \frac{i}{2}\int\left(
                          \overline{\psi_{\Lambda s}^{(-)}(q',\lambda)}\hat\gamma_{(a}\Gamma_{b)}
                          \psi_{\Lambda s}^{(-)}(q',\lambda) -
                          \overline{(\Gamma_{(a}\psi_{\Lambda s}^{(-)}(q',\lambda))}
                          \hat\gamma_{b)}\psi_{\Lambda s}^{(-)}(q',\lambda)
                         \right)\times \\ \nonumber
 & \rule{1cm}{0pt}\times d\mu(q')d\mu(\lambda)d\mu(\Lambda)d\mu(s).
\end{align}

The SEMT components (\ref{T00_vac}-\ref{TAB_vac}) and (\ref{T00_vac_sp}-\ref{Tab_vac_sp}) do not depend on local coordinates. These components are expressed by the operators $\lambda$-representation of the Lie algebra and solutions of the differential equation in the K-orbit. 

The work has been partially supported by Ministry of Education and Science of the Russian Federation under the analytical program No 2.1.1/12999, the Russian Federal program “Kadry” under the contracts No 02.740.11.0238; P691; P789.


\end{document}